\documentclass{SciPost}

\hypersetup{
    colorlinks,
    linkcolor={red!50!black},
    citecolor={blue!50!black},
    urlcolor={blue!80!black}
}

\usepackage{comment}
\usepackage[bitstream-charter]{mathdesign}
\usepackage{soul}
\usepackage{siunitx}

\DeclareSymbolFont{usualmathcal}{OMS}{cmsy}{m}{n}
\DeclareSymbolFontAlphabet{\mathcal}{usualmathcal}

\fancypagestyle{SPstyle}{
\fancyhf{}
\lhead{\colorbox{scipostblue}{\bf \color{white} ~SciPost Physics }}
\rhead{{\bf \color{scipostdeepblue} ~Submission }}

\fancyfoot[C]{\textbf{\thepage}}
}

\definecolor{DarkBlue}{rgb}{0,0,0.80}
\definecolor{DarkRed}{rgb}{0.80,0,0}
\definecolor{Purple}{rgb}{0.55,0,0.55}
\definecolor{Purple}{rgb}{0,0,0.8}

\usepackage{braket}

\begin{document}
\pagestyle{SPstyle}

\begin{center}{\Large \textbf{\color{scipostdeepblue}{
Coherent control of spinmons\\
}}}\end{center}

\begin{center}\textbf{
Johanne Bratland Tjernshaugen\textsuperscript{1$\star$},
Florinda {Viñas Boström}\textsuperscript{2},
Jeroen Danon\textsuperscript{1},\\
Jacob Linder\textsuperscript{1},
Karsten Flensberg\textsuperscript{2} and
Antonio L.~R.~Manesco\textsuperscript{2$\dagger$}
}\end{center}

\begin{center}
{\bf 1} Center for Quantum Spintronics, Department of Physics, Norwegian \\ University of Science and Technology, NO-7491 Trondheim, Norway
\\
{\bf 2} Center for Quantum Devices, Niels Bohr Institute, University of Copenhagen, DK-2100 Copenhagen, Denmark
\\[\baselineskip]
$\star$ \href{mailto:email1}{\small johanne.b.tjernshaugen@ntnu.no}\,,\quad
$\dagger$ \href{mailto:email2}{\small spinmon@antoniomanesco.org}
\end{center}

\section*{\color{scipostdeepblue}{Abstract}}
\textbf{\boldmath{
In this work, we explore a hybrid qubit device where quantum information is encoded by quantum states of a transmon entangled with the spin of a trapped Andreev quasiparticle.
We term such devices \textit{spinmons}.
We lift the spinmon Kramers degeneracy by introducing a Zeeman field and develop two routes for full qubit control via electrostatic gates and an AC flux drive, providing multiple directions for experimental implementations.
Finally, we compute coherence times, verify the qubit robustness against flux, charge and gate voltage noise sources, and discuss the role of nuclear spin noise. 
}}

% \vspace{\baselineskip}

%%%%%%%%%% BLOCK: Copyright information
% This block will be filled during the proof stage, and finilized just before publication.
% It exists here only as a placeholder, and should not be modified by authors.
% \noindent\textcolor{white!90!black}{%
% \fbox{\parbox{0.975\linewidth}{%
% \textcolor{white!40!black}{\begin{tabular}{lr}%
%   \begin{minipage}{0.6\textwidth}%
%     {\small Copyright attribution to authors. \newline
%     This work is a submission to SciPost Physics. \newline
%     License information to appear upon publication. \newline
%     Publication information to appear upon publication.}
%   \end{minipage} & \begin{minipage}{0.4\textwidth}
%     {\small Received Date \newline Accepted Date \newline Published Date}%
%   \end{minipage}
% \end{tabular}}
% }}
% }
%%%%%%%%%% BLOCK: Copyright information

% For convenience during refereeing we turn on line numbers:
%\linenumbers

% \vspace{10pt}
% \noindent\rule{\textwidth}{1pt}
% \tableofcontents
% \noindent\rule{\textwidth}{1pt}
% \vspace{10pt}

% Write your article contents here, starting from first \section.
% An example structure is given below.

\section{Introduction}
\label{sec:intro}

The search for superconducting qubits with longer lifetimes has inspired the
exploration of different circuit designs.
In particular, superconducting qubit Hamiltonians with more than one degree of freedom have received increasing attention~\cite{PRXQuantum.2.030101, Danon_2021}, since they tend to exhibit increased protection against decoherence.
For this reason, recent efforts have focused on developing multimode qubits, in which the logic-state wave functions reside in disjoint parts of the Hilbert space~\cite{Chakraborty_2025, PRXQuantum.2.030101, Danon_2021}.
Examples include the bifluxon \cite{kalashnikov2020bifluxon}, the $0-\pi$ qubit~\cite{brooks2013protected, groszkowski2018coherence}, Josephson Rhombus chains~\cite{Ioffe_2002, bell2014protected}, $\cos 2\varphi$ qubits~\cite{smith2020superconducting, roverch2026experimentalrealizationcos2varphitransmon}, and cat qubits~\cite{Cochrane_1999, Mirrahimi_2014}. 

Another recent direction for qubit protection is combining microscopic and macroscopic quantum degrees of freedom.
In the context of superconducting elements, a natural way to entangle these degrees of freedom is by using Andreev quantum circuits, both at even~\cite{caceres2026ferbonoiseresilientqubit,Matute_Ca_adas_2024} and odd parity~\cite{kurilovich2026andreev, manesco2026looplessmultiterminalquantumcircuits}.
Mesoscopic Josephson junctions with spin-orbit coupling can host an odd-parity state where the spin of the excited quasiparticle acts as a qubit~\cite{PhysRevLett.90.226806, Padurariu_2010, Park_2017, Hays_2020, Hays_2021, bargerbos2023spectroscopy, pita2023direct, Pita_Vidal_2024}, which is known as an Andreev spin qubit.
The Andreev spin qubit is prone to relaxation, but by shunting the Josephson element with a capacitor, the spins live in disjoint parts of the superconducting phase space, which yields protection against this source of decoherence~\cite{kurilovich2026andreev}.
We term these capacitively shunted Andreev spin qubits \emph{spinmons}.

In the odd-parity state, spin-orbit coupling breaks spin-degeneracy, resulting in the spin-dependent energy-phase relation typical of an Andreev spin qubit~\cite{PhysRevLett.90.226806,Padurariu_2010,PhysRevB.109.155164}:
\begin{equation}
    U_\text{ASQ}(V_g,\varphi)=E_{0}(V_g)\cos\varphi + E_\text{SO}(V_g)\vec{n}_\text{SO}(V_g)\cdot\vec{\sigma}\sin\varphi~,
    \label{eq:asq-hamiltonian}
\end{equation}
where $E_0$ is the spin-independent part of the potential, $E_\text{SO}$ is the amplitude of the phase-dependent spin-splitting, $\varphi$ is the phase drop across the Josephson junction, $V_g$ is a gate voltage, and $\vec{\sigma} = (\sigma_x, \sigma_y, \sigma_z)$ are Pauli matrices acting in the spin degree of freedom. $\vec{n}_\text{SO}$ is the direction of the spin-orbit field.
For a single-channel Josephson junction at odd parity, typically $E_0 > 0$~\cite{bulaevskii1977superconducting, van2006supercurrent}.
Shunting such a Josephson element with a capacitor results in two spin-dependent potential wells localized at $\pi \pm \varphi_0$, and in the heavy transmon limit, the spin states live at the minima of their respective wells~\cite{kurilovich2026andreev, manesco2026looplessmultiterminalquantumcircuits}.
The separation of the wells by $2\varphi_0$ protects this device from spin-flipping caused by a nuclear spin bath, via the Franck-Condon blockade~\cite{kurilovich2026andreev}.
Since nuclear spins are a limiting factor for lifetimes of Andreev spin qubits in InAs platforms~\cite{Lu_2025}, capacitive shunting offers a route for improving $T_1$ in these devices.

\begin{figure*}%[!h]
    \centering
    \includegraphics[width=0.8\textwidth]{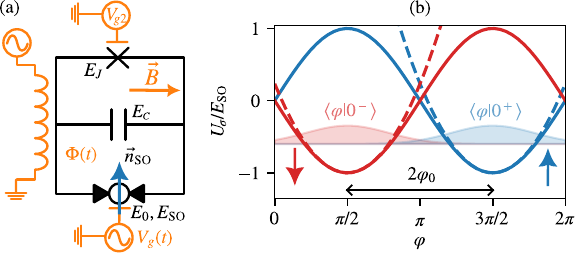}
    \caption{(a) Spinmon with flux and gate voltage control. 
    The purpose of $V_{g2}$ is to tune the critical current of the corresponding semiconductor-based Josephson junction, and thus modify $E_J$.
    The voltage $V_g$ is used to induce spin rotations via electron dipole spin resonance, effectively by changing the direction of $\vec{n}_\text{SO}$.
    The $\vec{B}$-field splits the computational states.
    The AC drive line to the left in the figure is used to induce a magnetic flux $\Phi$, which lifts the degeneracy of the potential energy wells in the spinmon.
    (b) Potential landscape of the spinmon in the spin-$\uparrow$ (blue) and spin-$\downarrow$ (red) idle state, where the spin quantization axis is $\vec{n}_\text{SO}$ at $\Phi=0$ and $E_0=E_J(V_{g2})$.
    The dashed line shows the harmonic approximations $U_\sigma^H$ to the potentials, with corresponding ground state wave functions $\ket{0^\sigma}$ projected onto the phase space.}
    \label{fig:setup}
\end{figure*}

In this work, we extend the theoretical proposal of Ref.~\cite{kurilovich2026andreev} of a spinmon by describing practical routes for coherent manipulation while showing it remains robust against noise.
Our proposed setup, schematically shown in Fig.~\ref{fig:setup}(a), consists of adding a mesoscopic Josephson junction at even parity in parallel to a capacitively shunted Andreev spin. The Josephson current of the even-parity junction is tunable via a gate voltage. This represents an advantage compared to the simpler spinmon in Ref. \cite{kurilovich2026andreev}, since it allows for electrostatically tuning the Franck-Condon suppression of spin flips, including tuning to the point with maximal protection $\varphi_0=\pi/2$, without changing any properties of the odd-parity junction. Additionally, Zeeman splitting makes our spinmon less susceptible to the remaining decoherence mechanism from nuclear spins discussed in Ref. \cite{kurilovich2026andreev}.
The introduction of the loop enables controlled qubit rotations with flux, while a gate voltage on the odd-parity junction provides full electrical control of the qubit rotations.
.

\section{Model}

We propose full control of a spinmon, schematically shown in Fig.~\ref{fig:setup}(a), exploring both its electrostatic and flux tunability.
The main concept behind our proposal
is connecting the odd-parity Josephson junction in parallel to an even-parity gate-tunable Josephson junction.
Neglecting the self-inductance of the loop, the resulting Hamiltonian of the circuit is 
\begin{equation}\label{eq:full Hamiltonian}
    H = -4E_c\partial_\varphi^2+ U_\text{ASQ}(V_g, \varphi)- E_{J}(V_{g2})\cos \varphi - \vec{B}\cdot \vec{\sigma}, 
\end{equation}
where $E_c$ is the charging energy of the capacitor, $E_J(V_{g2})$ is the Josephson energy of the even-parity junction, tunable via the gate voltage $V_{g2}$, and $\vec{B}$ is an external Zeeman field.
Our qubit states $\ket{0_q}, \ket{1_q}$ are taken as the ground-state and first excited state of $H$.
We fix the quantization axis $\vec{n}_{\rm SO}=\hat{z}$  when the qubit is idle, and define the spin-dependent potential $U_\sigma(\varphi) = \delta E\cos\varphi+E_{\rm SO}\sigma\sin\varphi$ with $\delta E = E_0 - E_J(V_{g2})$.

To proceed analytically, we focus on the heavy spinmon limit, $E_C \ll E_{0} \sim E_\text{SO} \sim E_J$, where we can approximate the spin-dependent potentials as harmonic potentials, $U_{\sigma}^H$,
\begin{equation}\label{eq:fluxH}
\begin{split}
    H = -4E_c\partial_{\varphi}^2 + U^H - \vec{B}\cdot \vec{\sigma}, \quad U_{\sigma}^H=- \tilde{E}_J\left\lbrace1-\frac{1}{2}\left[\varphi -(\pi - \sigma \varphi_0)\right]^2\right\rbrace
\end{split}
\end{equation}
with
\begin{equation}
    \label{eq:idle_params}
    \tilde{E}_J = \sqrt{\delta E^2 + E_\text{SO}^2}~,\quad \varphi_0 = \arctan\left(\frac{E_\text{SO}}{\delta E}\right)~.
\end{equation}
We show the spin-dependent harmonic potential $U_\sigma^H$ at $\delta E=0$ in Fig.~\ref{fig:setup}(b) (dashed curves) together with the exact potential $U_\sigma$ (solid curves).
In the absence of $B$, the two lowest states of the spinmon are the Kramers pair illustrated in Fig.~\ref{fig:setup}(b).
We denote these states as $\ket{0^{\tau}\sigma} \equiv\ket{0^{\tau}}\otimes\ket{\sigma}$, where $\ket{0^{\tau}}$ denotes the lowest eigenstate in the harmonic potential well centered at $\pi + \tau \varphi_0$, $\tau=\pm$, and $\ket{\sigma}$ is its spin state.
A Zeeman field $\vec{B}\parallel \hat{x}$ mixes the two spin states and breaks the Kramers degeneracy.
We assume that $|\vec{B}| \ll \omega_p$, with $\omega_p = \sqrt{8E_c\tilde E_J}/\hbar$ the Josephson plasma frequency, which allows us to project the Zeeman term on the low-energy subspace $\{\ket{0^+\uparrow},\ket{0^-\downarrow}\}$.
This yields the symmetric and antisymmetric states $\ket{0} = (\ket{0^+\uparrow}-\ket{0^-\downarrow})/\sqrt{2}$ and $\ket{1} = (\ket{0^+\uparrow}+\ket{0^-\downarrow})/\sqrt{2}$ as the lowest eigenstates, so that $\ket{0}\simeq\ket{0_q}$ and $\ket{1}\simeq\ket{1_q}$, with a splitting
\begin{equation}
    \label{eq:splitting}
    \hbar\omega_q = 2Be^{-\xi_0 / 2}~,\quad \xi_0 = \left(\frac{\varphi_0}{\varphi_c}\right)^2~,\quad
    \varphi_c = \sqrt{\frac{4E_C}{\hbar\omega_p}}
\end{equation}
where $B=|\vec{B}|$.
Spin-mixing terms $\sim \sigma_{x, y}$ projected onto the qubit basis are proportional to the Franck-Condon factor $\text{e}^{-\xi_0/2} = \braket{0^+ | 0^-}$, reflecting the exponentially small overlap between the spin states in phase space in the heavy transmon limit. 

\begin{figure}[h]
    \centering
    \includegraphics[width=\textwidth]{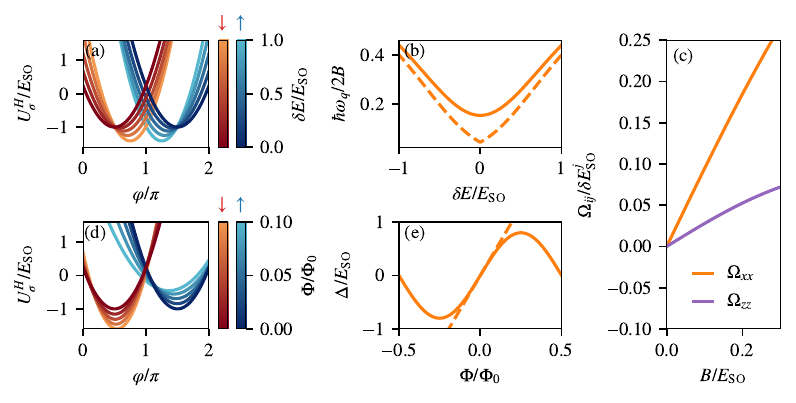}
    \caption{Coherent control of the spinmon.
    (a) The spin-dependent potentials as a function of $\varphi$ for varying $\delta E$.
    The shade of blue(red) shows how the spin-up(down) potential evolves as a function of $\delta E = E_0-E_J(V_{g2})$, see the color bar.
    (b) The qubit frequency as a function of $\delta E/E_{\rm SO}$.
    The solid curve is calculated numerically by diagonalizing the full Hamiltonian in Eq.~\eqref{eq:full Hamiltonian}, and the dashed curve is calculated analytically in the harmonic oscillator approximation.
    (c) The combination of Zeeman splitting and a gate voltage permits an EDSR driving with strength $\Omega_{xx}$, and also renormalizes the qubit splitting via diagonal elements $\Omega_{zz}$ in the qubit basis. All finite matrix elements are shown.
    (d) The spin-dependent potentials as a function of $\varphi$ for varying external flux $\Phi$.
    (e) A small flux gives a qubit-flipping term in the effective qubit Hamiltonian with strength $\Delta$.
    The dashed line is linearized in flux.
    We have set $E_0=E_\text{SO}=h\times \qty{500}{\mega\hertz}$ and $B=E_c=h\times\qty{40}{\mega\hertz}$. 
    In (a)-(c) the flux is fixed at $\Phi=0$ and in (c)-(e) we use $\delta E=0$. }
    \label{fig:fluxcontrol}
\end{figure}

For the remainder of the manuscript, we set $E_0=E_\text{SO}=h\times\qty{500}{\mega\hertz}$, comparable to previous experimental measurements~\cite{Hays_2021, bargerbos2023spectroscopy, Lu_2025}; $E_c=h\times\qty{40}{\mega\hertz}$, comparable to state-of-the-art experiments~\cite{kurilovich2025highfrequencyreadoutfreetransmon}; and set $B=E_c$.
Since $V_{g2}$ controls $\delta E = E_0 - E_J(V_{g2})$, it is possible to tune the distance between the spin-dependent wells while keeping $E_0$ and $E_{SO}$ unchanged, as shown in Fig.~\ref{fig:fluxcontrol}(a).
This results in an electrostatic control of the qubit splitting, as shown in Fig.~\ref{fig:fluxcontrol}(b). 
This control of the qubit splitting enables initialization and readout via a readout resonator. Since the wave function overlap depends on $\delta E$, as shown in Fig. \ref{fig:fluxcontrol}(b), we can maximize the qubit protection against spin flipping by setting $\delta E = 0$, which we define as the idle configuration of the spinmon.
Notably, if $E_0 > 0$, we can achieve the same idle state by threading half a flux quantum through the Josephson loop.
The discrepancy between the analytical and numerical results for the qubit splitting in Fig.~\ref{fig:fluxcontrol}(b) arises because the qubit splitting is proportional to the overlap integral $\braket{0^+ | 0^-}$, whose main contributions come from the parts of the phase space where the difference between the harmonic and full potentials is maximal.
This explains why the discrepancy is greatest when the separation between the wells is maximized.

\section{Coherent control}

\subsection{Electric dipole spin resonance}
\label{sec:edsr}

Previous experimental works used electric dipole spin resonance (EDSR) to flip Andreev spins via AC driving of a gate voltage~\cite{pita2023direct}.
We propose the same strategy for Rabi driving of the spinmon with $V_g$.
We include the EDSR effects by considering the lowest-order corrections to the spin-dependent term in Eq. \eqref{eq:asq-hamiltonian} due to an oscillating electric field resulting from an AC drive of the gate voltage $V_g$.
To linear order in $V_g$, the Hamiltonian changes as
\begin{equation}
    H(V_g + \delta V_g) - H(V_g) = \delta V_g\left(\frac{\partial E_\text{SO}}{\partial V_g} \sigma_z + E_\text{SO}\frac{\partial \vec{n}_\text{SO}}{\partial V_g}\cdot \vec{\sigma}\right)\sin \varphi = \vec{\delta E_\text{SO}} \cdot \vec{\sigma} \sin \varphi~,
    \label{eq:gate-voltage}
\end{equation}
where we neglect the changes in $\delta E$ because the qubit Hamiltonian is insensitive to leading-order contributions in this parameter, as seen from the solid line in Fig.~\ref{fig:fluxcontrol}(b).
This spin-orbit modulation does not flip spins to lowest order, \emph{i.e.},  $\langle 0 |\sigma_j \sin\varphi|1\rangle=0$ for any $j$.
However, by using the exact eigenstates of the Hamiltonian, including corrections from the Zeeman field, the matrix element becomes finite, as confirmed by numerical calculations.
The effective Hamiltonian is
\begin{equation}
    H_q = \frac12\hbar\omega_q\tau_z + \sum_{i,j} {\Omega}_{ij} {\tau}_i~,\quad \Omega_{ij} = \frac{1}{2}\mathrm{tr}\left(\tau_i \sum_{l, l'\in \{0, 1\}} \ket{l_q}\bra{l_q} \delta E_\text{SO}^j \sigma_j \sin \varphi \ket{l'_q}\bra{l'_
    q} \right)~,
    \label{eq:edsr_qubit}
\end{equation}
where $\tau_i$ are the Pauli matrices in the qubit subspace.
The $B$-dependence of the non-zero matrix elements $\Omega_{ij}$, as well as the renormalization of the qubit splitting induced by the presence of $\delta V_g$, are shown in Fig.~\ref{fig:fluxcontrol}(c).
We observe that both $\Omega_{xx}$ and $\Omega_{zz}$ are nonzero.
An oscillating term $\sim \Omega_{xx}$ drives Rabi oscillations, therefore allowing full electrostatic control of the qubit.

While, as we argued above, Rabi oscillations are generically achieved via EDSR, the exact dependence of the spin-orbit vector on gate voltages is sensitive to microscopic details of the electrostatic environment and the disorder landscape across the device \cite{han2023variable, bargerbos2023spectroscopy}.
Therefore, predicting its gate dependence is challenging even with detailed microscopic modeling of the device.

\subsection{Flux drive}

An alternative for coherent control is the use of an AC flux line that can induce oscillations of the flux $\Phi$, see Fig.~\ref{fig:setup}(a).
We take linear corrections to the applied flux, which, in the limit where the shunt capacitance dominates the charging term, changes the potential of the circuit Hamiltonian as (see Appendix~\ref{app:circuit_hamiltonian} for a full derivation):
\begin{equation}
    \delta U_\sigma(\varphi)
    \equiv U_\sigma(\varphi;\phi)-U_\sigma(\varphi;0)
    =-E_J\!\left[\cos(\varphi-\phi)-\cos\varphi\right]
    =-\phi E_J\sin\varphi+\mathcal O(\phi^2),
    \label{eq:flux_effect}
\end{equation}
where $\phi=2\pi \Phi / \Phi_0$, $\Phi$ is the magnetic flux through the loop, and $\Phi_0$ is the superconducting flux quantum.
We choose the ASQ phase as the circuit coordinate, so that the external phase appears explicitly in the conventional JJ potential.
The time-reversal symmetry breaking caused by the flux lifts the Kramers degeneracy of the wells.
As shown in Fig.~\ref{fig:fluxcontrol}(d), at small $\phi$ the two wells split linearly while the distance between the wells remains unchanged.
Projecting $H(\phi)$ onto the qubit basis, we obtain
\begin{equation}
    H_q(\phi) = \frac{1}{2}\hbar \omega_q(V_{g2})\tau_z+\Delta(\phi)\tau_x~,
    \label{eq:flux_qubit}
\end{equation}
where $\Delta(\phi)$ is the qubit flipping matrix element resulting from projecting Eq.~\eqref{eq:flux_effect} onto the qubit basis.
In the numerical calculations, we include the full potential dependence on $\phi$, \emph{i.e.} beyond linear approximation.
We note that the flux also gives rise to a small change in the qubit frequency when using the exact eigenstates and all orders of $\phi$, which we estimated numerically to be negligible compared to $\Delta$.
The flux dependence of $\Delta$ is shown in Fig. \ref{fig:fluxcontrol}(e).
Its slope at $\phi=0$ determines the transverse coupling, and hence the Rabi frequency, produced by a weak resonant AC flux drive.
Therefore, flux drive and gate-voltage control of the auxiliary Josephson junction allow full control of the qubit.

\section{Decoherence}

\begin{figure}
    \centering
    \includegraphics[width=\textwidth]{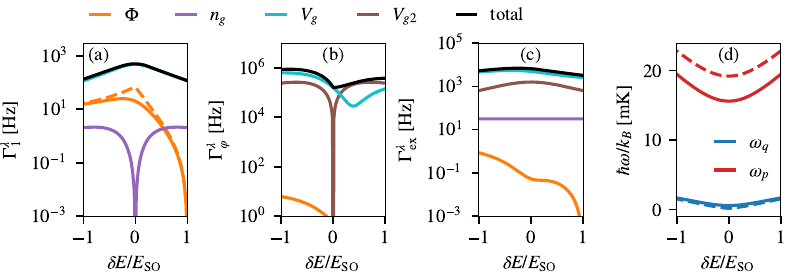}
    \caption{
    Decoherence of the spinmon.
    The solid lines are calculated numerically by diagonalizing the full Hamiltonian in Eq. \eqref{eq:full Hamiltonian}, and the dashed lines are calculated analytically in the harmonic oscillator approximation.
    (a) Relaxation rates for flux noise ($\lambda=\Phi$), charge noise on the capacitor ($\lambda=n_g$), and gate voltage noise ($\lambda=V_g, V_{g2}$). The black line is the sum of the four rates.
    (b) Dephasing rates.
    (c) Excitation rate out of the computational subspace.
    (d) Qubit frequency $\omega_q$ and plasma frequency $\omega_p$ in units of temperature.
    Numerically, $\omega_p$ is the frequency difference between the first and second excited states.
    We have set $E_0=E_\text{SO}=h\times \qty{500}{\mega\hertz}$ and $B=E_c=h\times\qty{40}{\mega\hertz}$.
    }
    \label{fig:decoherence}
\end{figure}

Introducing flux and gate voltage control of the spinmon opens additional channels for dephasing and relaxation. First, as for all qubits with loops, the spinmon becomes sensitive to flux noise. Since $\Delta(\Phi)\propto \Phi$ for small $\Phi$, flux noise causes qubit flips, which we estimate from Fermi's golden rule~\cite{krantz2019quantum},
\begin{equation}
    \Gamma_1^{\lambda}=\frac{1}{\hbar^2} |\bra{0_q}
    \partial_\lambda H
    \ket{1_q}|^2 S_{\lambda}(\omega_q),
\end{equation}
where flux noise ($\lambda=\Phi$) typically has a $1/f$ spectral density $S_\lambda(\omega_q) = 2\pi A_\lambda^2/\omega_q$ and $A_\Phi =$ $10^{-6}\Phi_0$ \cite{hutchings2017tunable}.
With the parameters used throughout the manuscript, we obtain a rate below $\qty{100}{\hertz}$, as shown in Fig. \ref{fig:decoherence}(a). The dephasing rate from $1/f$ noise is estimated by
\begin{equation}
    \Gamma_\varphi^\lambda = A_\lambda |\partial_\lambda \omega_q|\sqrt{2|\mathrm{ln}(\omega_{\rm ir}t)|}
\end{equation}
where $\omega_{\rm ir}$ is an infrared cutoff and $t$ is the time the qubit is exposed to the noise, and a typical value of $\omega_{\rm ir}t=2\pi \times 10^{-5}$.
The dephasing rate from flux noise is zero. \ref{fig:decoherence}(b).
The relaxation and dephasing rates from charge noise ($\lambda=n_g$) is calculated by inserting the offset charge $n_g$ of the shunting capacitor by $\partial_\varphi \rightarrow \partial_\varphi+\text{i}n_g$ in Eq. \eqref{eq:full Hamiltonian}.
The relaxation and dephasing rates from both $1/f$, shown in Fig. \ref{fig:decoherence}(a), and ohmic charge noise (not shown) at $n_g=0$ are negligible compared to the other noise sources.
We assumed that the amplitude of the for $1/f$ charge noise is $A_{n_g}=10^{-4}e$\cite{zorin1996background}.
The spinmon's insensitivity to charge noise is reasonable considering that the transmon is designed to be insensitive to charge noise. 
Flux and charge noise also give rise to excitations out of the computational space.
The excitation rate out of the computational subspace is given by 
\begin{equation}
    \Gamma_\text{ex}^\lambda = \frac{1}{\hbar^2}\sum_{i=0,1}\sum_{m\geq 2} |\bra{i_q}\partial_\lambda H \ket{m_q}|^2S_\lambda(\omega_m-\omega_i),
\end{equation}
where $\omega_{i,m}$ are the eigenfrequencies of $\ket{i_q,m_q}$.
The excitation rates at $T=\qty{10}{\milli\kelvin}$ are shown in Fig. \ref{fig:decoherence}(c).
They are both below $\qty{100}{\hertz}$, demonstrating that excitations out of the computational subspace from flux and charge noise are not a bottleneck for the spinmon lifetimes.

The electrostatic environment required for gate control and for stabilizing the odd-parity state introduces another source of charge noise that is effectively manifested as gate-voltage noise.
The gate-sensitivity of the Hamiltonian to linear order in $V_g$ and $V_{g2}$ is 
\begin{equation}
    \partial_{V_g}H = (\partial_{V_g} E_0) \cos(\varphi) + \partial_{V_g}(E_\mathrm{SO}\vec{n}_\mathrm{SO})\cdot \vec{\sigma} %\sum_i (\partial_{V_g}E_\mathrm{SO}^i)\sigma_i
    \sin(\varphi)~,\quad
    \partial_{V_{g2}}H = -(\partial_{V_g} E_J) \cos(\varphi)
\end{equation}
We estimated the lever arms $\lambda_0 = \partial_{V_g}E_0/h$ and $\lambda_\mathrm{SO} = \partial_{V_g} {E}_\mathrm{SO}n_\mathrm{SO}^i/h$ from the publicly available dataset of Ref.~\cite{bargerbos2023spectroscopy} to be both on the order of $\qty{10}{\giga\hertz/\volt}$. The data analysis is included in our Zenodo repository~\cite{tjernshaugen_2026_20082243}.
We also found that the lever arm for gate-tunable superconducting qubits is typically $\sim\qty{10}{\giga\hertz/\volt}$ \cite{purkayastha2026transmon, isakov2026observation}, which we adopt as the value for $\lambda_J = \partial_{V_{g2}}E_J/h$.
Finally, following previous works suggesting an $1/f$ spectral density with $A_{V_g} \sim \qty{10}{\micro \volt}$~\cite{PhysRevLett.120.100502, Zatelli_2024}, we computed the relaxation rate, dephasing rate, and excitation rate shown in Fig. \ref{fig:decoherence}(a)-(c).
Noise in $\delta E$ causes no relaxation, and since the qubit frequency is minimal at the idle point $\delta E=0$, dephasing is also suppressed, as shown in Fig.~\ref{fig:decoherence} (b).
Away from this point, $\delta E$-noise contributes to a dephasing rate $\sim \qty{100}{\kilo\hertz}$, which may be a bottleneck for single-qubit gate fidelities.
Noise in the spin-orbit field, on the other hand, contributes to both relaxation and dephasing, and is the bottleneck for qubit lifetimes, leading to $\Gamma_1\sim\qty{1}{\kilo \hertz}$ and $\Gamma_{\phi}\sim\qty{100}{\kilo\hertz}$.
The asymmetric dependence of the dephasing rate with $\delta E$ is due to the combined symmetric contribution from noise in $E_0$ and the antisymmetric contribution from $E_{\mathrm{SO}}$.
These results encourage engineering the electrostatic environment to minimize the effects of charge noise on $E_{\mathrm{SO}}$ or exploring multi-dot architectures to suppress these noise effects. 

The spinmon is protected from nuclear spin noise perpendicular to the spin-orbit axis.
However, the Franck-Condon blockade still leaves an unprotected decoherence channel due to nuclear spin noise when the exchange field is parallel to the spin-orbit axis.
While this mechanism causes dephasing in Andreev spin qubits, limiting lifetimes to tens of nanoseconds~\cite{Hays_2021,bargerbos2023spectroscopy,tjernshaugen_2026_20082243}, in a spinmon it causes qubit relaxation.
Because nuclear spin dynamics are much slower than qubit precession, the spectral power of this noise channel at the qubit frequency is negligible.
Together with the limited energy available in nuclear spin flips, this motivates the expectation of weak direct nuclear-induced relaxation away from degeneracies.
Thus, we expect spinmons to be much less sensitive to nuclear spin noise than standard Andreev spin qubits.
Nevertheless, their lifetimes are still expected to benefit from recent progress in isotopically purifiable materials, such as planar germanium ~\cite{Lakic_2025, PRXQuantum.5.030357, fabris2026granularaluminuminducedsuperconductivity}, graphene~\cite{jung2025tunneling}, and carbon nanotubes~\cite{Riechert_2025}.

Since $\hbar\omega_p / k_B$ is on the order of tens of millikelvin and $\hbar\omega_q/k_B$ is even smaller, as shown in Fig. \ref{fig:decoherence}(d), thermal excitations become a bottleneck for the spinmon lifetimes and initialization.
It is possible to achieve the strict temperature requirements using sideband cooling, previously used for heavy fluxonium with $\omega_q = \qty{1.8}{MHz}$ \cite{najerasantos_prx_24}.
The energy scales of the spinmon are limited by the small magnitude of $E_\text{SO}$, so electrostatic engineering of the confinement potential to increase $E_\text{SO}$ would be an interesting avenue for decreasing the spinmon's sensitivity to thermal excitations.

\section{Outlook}

We propose full control of a spinmon via a plunger gate, EDSR, and flux control, but a universal set of quantum gates still requires two-qubit coupling.
This can be achieved via inductive coupling between two qubits of the type $\sim I_1I_2$ in the Hamiltonian, which in the presence of flux $\Phi$ results in a $\tau_x^{(1)}\tau_x^{(2)}$ two-qubit coupling in the qubit basis.
The flux-dependence is advantageous since it allows for a tunable coupler, similarly to long-range coupling of Andreev spin qubits~\cite{Pita_Vidal_2024}.
We leave a systematic construction of two-qubit gates and estimation of gate fidelities for future work, since such an analysis would require an investigation of the dependence of the spin-orbit field $\vec{n}_\text{SO}$ on microscopic details of the device.

The introduction of a loop in the spinmon introduces flux noise as a new noise channel that reduces $T_1$.
Although this noise channel does not significantly decrease the qubit lifetime in the presence of the Zeeman field $B=h \times\qty{40}{\mega\hertz}$, the field-free realization of the spinmon is degenerate and its lifetimes are limited by $1/f$ flux noise.
A possible route for avoiding spin relaxation due to flux noise is to shunt the looped spinmon with an inductor.
This strategy has been proposed to suppress flux noise used for fluxonium~\cite{Bao_2022}, bifluxon~\cite{kalashnikov2020bifluxon}, and Andreev qubits~\cite{caceres2026ferbonoiseresilientqubit}.
Another strategy to mitigate flux noise is using gradiometric designs~\cite{Schwarz_2013}, as recently realized in flux qubits~\cite{krojer2024fast} and in a tunable transmon~\cite{fu2026fluxnoiseresilienttransmonqubitdoublyconnected}.
Alternatively, it is possible to explore the gate dependence of the ratio $E_0/E_\text{SO}$, as previously probed in Ref.~\cite{bargerbos2023spectroscopy}, to completely remove the loop.
The possibility of loopless Andreev spin devices in two-terminal devices would simplify recent proposals of multiterminal realizations~\cite{manesco2026looplessmultiterminalquantumcircuits} and make it compatible with nanowire setups.

Because misalignments to the external Zeeman field introduce additional flux through the loop, one could also explore proximitizing the semiconductor to a magnetic insulator.
This direction would also admit combining multiple Josephson junctions with various orientations.
This direction makes our proposal compatible with superconductor-ferromagnet-semiconductor structures~\cite{Liu_2019, Vaitiek_nas_2020, PhysRevB.105.L041304, Jiang_2025}.

\section{Conclusion}

We developed the theory for coherent control of a spinmon using three ingredients.
The first one is a gate-tunable SQUID, where one of the Josephson junctions is in the even-parity state, and the second one is at odd-parity.
The gate-tunability of the odd-parity junction allows Rabi oscillations via spin flipping resulting from electric dipole spin resonance (EDSR).
This happens when taking into account the second ingredient, which is a static Zeeman field.
Combined with the gate tunability of the even-parity junction, this also provides electric control of the qubit splitting.
The third ingredient is an AC flux line, which provides a tool for Rabi oscillations.
Thus, our investigation suggests three paths for coherent control, providing flexibility for experimental realizations.
Finally, we demonstrated the qubit protection by computing typical rates for common decoherence processes, and highlighted future directions for the development of this qubit platform.

\section*{Acknowledgments}
We thank William Lawrie, Valla Fatemi, Bernard van Heck and Stefano Bosco for useful discussions. JBT is grateful for the hospitality of QDev and the Niels Bohr Institute while carrying out part of the work. 

\paragraph{Author contributions}
KF initiated the project with input from JD, JBT, ALRM, and FVB.
JBT and FVB performed the analytical calculations, and JBT and ALRM performed the numerical simulations, with input from all authors.
JBT, JL, and ALRM wrote the manuscript with input from all authors.
ALRM, KF, and JL supervised the project.

\paragraph{Funding information}
FVB acknowledges funding from the European Union’s Horizon Europe research and innovation programme under the Marie Sk\l{}odowska-Curie grant agreement No 101204715.
ALRM acknowledges the funding from the European Research Council (Grant Agreement No.~856526).
JBT and JL were supported by the Research Council of Norway
through Grant No. 353894 and its Centres of Excellence
funding scheme Grant No. 262633 “QuSpin.”

\paragraph{Code and data availability}
The code and data used to prepare this manuscript are available in Zenodo \cite{tjernshaugen_2026_20082243}.

\begin{appendix}

\section{Circuit Hamiltonian for arbitrary flux}
\label{app:circuit_hamiltonian}

\subsection{Flux constraints and phase parametrization}

Let $\theta_0$ and $\theta_J$ denote the gauge-invariant phase
differences across the ASQ and conventional Josephson junction,
respectively. We assume that the flux line threads the loop formed by
the two Josephson-junction branches, while its flux linkage to the
shunt wiring is negligible. Neglecting the self-inductance of this
loop, flux quantization gives
\begin{equation}
    \theta_0(t)-\theta_J(t)
    =
    \phi(t),
    \qquad
    \phi(t)
    =
    \frac{2\pi\Phi(t)}{\Phi_0}.
    \label{eq:app_flux_constraint}
\end{equation}

We retain the freedom to distribute the externally imposed phase
between the two junction phases by introducing the parametrization
\begin{equation}
    \theta_0
    =
    \varphi+\alpha\phi(t),
    \qquad
    \theta_J
    =
    \varphi-(1-\alpha)\phi(t),
    \label{eq:app_phase_parametrization}
\end{equation}
where $\alpha$ is an arbitrary choice of phase coordinate. Different
values of $\alpha$ correspond to different representations of the
same physical circuit.
For definiteness, we define the shunt-capacitor coordinate using the ASQ branch, $\theta_s\equiv\theta_0$; using Eq.~\eqref{eq:app_phase_parametrization}, this gives $\theta_s=\varphi+\alpha\phi(t)$ for arbitrary $\alpha$.
Thus, the voltage in the shunt capacitor is
\begin{equation}
    V_s
    =
    \frac{\hbar}{2e}
    \left[
        \dot{\varphi} + \alpha\dot{\phi}(t)
    \right].
    \label{eq:app_shunt_voltage}
\end{equation}

Let $C_0$ and $C_J$ denote the effective intrinsic capacitances of the
ASQ and conventional-JJ branches, respectively, and let $C_s$ denote
the external shunt capacitance. The capacitive part of the Lagrangian is
\begin{equation}
    \mathcal L_C
    =
    \frac{1}{2}
    \left(
        \frac{\hbar}{2e}
    \right)^2
    \left\lbrace
        \left(C_s + C_0\right)\dot{\theta}_0^{\,2}
        +
        C_J\dot{\theta}_J^{\,2}
    \right\rbrace.
    \label{eq:app_capacitive_lagrangian}
\end{equation}
Substituting Eq.~\eqref{eq:app_phase_parametrization} gives
\begin{align}
    \mathcal L_C
    =
    \frac{1}{2}
    \left(
        \frac{\hbar}{2e}
    \right)^2
    \left\lbrace
        \left(C_s + C_0\right)\left[\dot{\varphi} + \alpha \dot{\phi}(t)\right]^{\,2}
        +
        C_J\left[\dot{\varphi}-(1-\alpha)\dot{\phi}(t)\right]^{\,2}
    \right\rbrace
    \label{eq:app_capacitive_lagrangian_expanded}
\end{align}

Since the momentum conjugate to $\varphi$ is
\begin{equation}
    p_\varphi
    =
    \frac{\partial\mathcal L_C}
    {\partial\dot{\varphi}}
    =
    \left(
        \frac{\hbar}{2e}
    \right)^2
    \left(
        C_\Sigma\dot{\varphi}
        +
        K_\alpha\dot{\phi}
    \right).
    \label{eq:app_canonical_momentum}
\end{equation}
where
\begin{equation}
    C_\Sigma
    =
    C_s+C_0+C_J,
    \qquad
    K_\alpha
    =
    C_{\Sigma}\alpha-C_J.
    \label{eq:app_CSigma_Kalpha}
\end{equation}

Writing $p_\varphi=\hbar\hat n$ and performing the Legendre
transformation gives, up to a term depending only on the externally
imposed flux,
\begin{equation}
    H_C
    =
    4E_C\hat n^2
    +
    \hbar\gamma_\alpha
    \dot{\phi}_{\mathrm{ext}}(t)\hat n
    +
    f(t),
    \label{eq:app_general_charging_hamiltonian}
\end{equation}
where
\begin{equation}
    E_C
    =
    \frac{e^2}{2C_\Sigma},
    \qquad
    \gamma_\alpha
    =
    \frac{
        C_J
    }{
        C_\Sigma
    } - \alpha.
    \label{eq:app_gamma_alpha}
\end{equation}
The function $f(t)$ contains no circuit operator and therefore does not
affect the circuit dynamics or transition matrix elements. It may be
omitted from the reduced circuit Hamiltonian.

The phase convention used in the main text corresponds to $\alpha = 0$.
In this case, $\gamma_\alpha=\gamma_0$ where
\begin{equation}
    \gamma_0
    =
    \frac{C_J}{C_\Sigma}.
    \label{eq:app_gamma_zero}
\end{equation}
Since the total capacitance is dominated by the external shunt,
$C_J/C_\Sigma\ll1$, the derivative coupling is expected to provide a
small correction. It is nevertheless retained below in order to
demonstrate gauge invariance exactly.

\subsection{Complete flux coupling in the main-text phase convention}

For $\alpha=0$, the complete Hamiltonian becomes
\begin{align}
    H(t)
    =
    4E_C\hat n^2
    +
    E_0\cos\varphi
    +
    E_{\mathrm{SO}}\sigma_z\sin\varphi
    -
    E_J
    \cos\left[
        \varphi-\phi(t)
    \right]
    -
    B\sigma_x
    +
    \hbar\frac{C_J}{C_\Sigma}
    \dot{\phi}(t)\hat n.
    \label{eq:app_alpha_zero_hamiltonian}
\end{align}
Thus, its difference from the static Hamiltonian of Eq.~\ref{eq:full Hamiltonian} for a small external flux 
in linear order is
\begin{equation}
    \delta H(t)
    =
    \phi(t)V_0^{\mathrm{pot}}
    +
    \hbar\gamma_0
    \dot{\phi}(t)\hat n,
    \label{eq:app_complete_perturbation_A}
\end{equation}
where
\begin{equation}
    V_0^{\mathrm{pot}}
    =
    -E_J\sin\theta~.
    \label{eq:app_VA_gamma0}
\end{equation}

We can change the choice of $\alpha$ in the time-dependent Hamiltonian via a translation operator $\mathcal U(t) = e^{i \alpha  \phi(t)\hat n}$.
Thus, requiring covariance of the time-dependent Schrodinger equation gives
\begin{equation}
    H_{\alpha}(t)
    =
    \mathcal U(t)H(t)
    \mathcal U^\dagger(t)
    +
    i\hbar
    \dot{\mathcal U}(t)
    \mathcal U^\dagger(t).
    \label{eq:app_time_dependent_transformation}
\end{equation}
And since
\begin{equation}
    i\hbar
    \dot{\mathcal U}(t)
    \mathcal U^\dagger(t)
    =
    -\hbar
    \alpha \dot{\phi}(t)\hat n~,
\end{equation}
the complete linear perturbation in the transformed representation is
\begin{equation}
    \delta H_{\alpha}(t)
    =
    \phi(t)V_{\alpha}^{\mathrm{pot}}
    +
    \hbar\gamma_{\alpha}
    \dot{\phi}(t)\hat n,
    \label{eq:app_complete_perturbation_J}
\end{equation}
where
\begin{equation}
    V_{\alpha}^{\mathrm{pot}}
    =
    (1-\alpha) V_0^{\mathrm{pot}}
    + \alpha
    \left(E_{\mathrm{SO}}\sigma_z\cos\varphi -E_0\sin\varphi\right).
    \label{eq:app_VJ}
\end{equation}

\subsection{Gauge invariance of the transition matrix element}

To verify explicitly that the transition matrix element is independent
of the phase convention, we note that
\begin{equation}
    V_\alpha^{\rm pot}-V_0^{\rm pot}
    =
    \alpha\,\partial_\varphi H_0
    =
    i\alpha[\hat n,H_0],
    \label{eq:app_potential_difference}
\end{equation}
where $H_0$ is the zero-flux Hamiltonian. Hence, for
$H_0|j_q\rangle=E_j|j_q\rangle$ and
$\hbar\omega_q=E_1-E_0$,
\begin{equation}
    \langle0_q|
    V_\alpha^{\rm pot}-V_0^{\rm pot}
    |1_q\rangle
    =
    i\alpha\hbar\omega_q
    \langle0_q|\hat n|1_q\rangle~,
\end{equation}
and zero for the diagonal matrix elements.

For a Fourier component $\phi(t)\propto e^{i\omega t}$, the complete
linear coupling operator is
\begin{equation}
    V_\alpha^{\rm eff}(\omega)
    =
    V_\alpha^{\rm pot}
    +
    i\hbar\omega
    \gamma_{\alpha}\hat n .
\end{equation}
It therefore follows that
\begin{equation}
    \langle0_q|
    V_\alpha^{\rm eff}(\omega)-V_0^{\rm eff}(\omega)
    |1_q\rangle
    =
    i\alpha\hbar(\omega_q-\omega)
    \langle0_q|\hat n|1_q\rangle .
\end{equation}
For a resonant transition, $\omega=\omega_q$, the right-hand side
vanishes. The transition matrix element, and hence the weak-drive
Rabi frequency and flux-noise relaxation rate, are therefore
independent of the phase convention $\alpha$.

In the $\alpha=0$ convention used in the main text, the remaining
charge-like flux-dependent coupling is proportional to $C_J/C_\Sigma$.
Since this additional term would only be sizable when the flux oscillation frequency $\hbar\omega \sim E_J C_\Sigma / C_J \sim \qty{1}{\tera\hertz}$, we neglect this term and use $V_0^{\rm pot}=-E_J\sin\varphi$ as the flux-coupling operator.

\end{appendix}

\bibliography{references}

@article{kurilovich2026andreev,
  title         = {{Andreev spin qubit protected by Franck-Condon blockade}},
  author        = {Kurilovich, PD and Vakhtel, T and Connolly, T and B{\o}ttcher, CGL and Van Heck, B},
  journal       = {Physical Review B},
  volume        = {113},
  number        = {11},
  pages         = {115429},
  year          = {2026},
  publisher     = {APS},
  doi           = {10.1103/l478-yp15}
}

@article{najerasantos_prx_24,
  title         = {{High-Sensitivity ac-Charge Detection with a MHz-Frequency Fluxonium Qubit}},
  author        = {Najera-Santos, B.-L. and Rousseau, R. and Gerashchenko, K. and Patange, H. and Riva, A. and Villiers, M. and Briant, T. and Cohadon, P.-F. and Heidmann, A. and Palomo, J. and Rosticher, M. and le Sueur, H. and Sarlette, A. and Smith, W. C. and Leghtas, Z. and Flurin, E. and Jacqmin, T. and Del\'eglise, S.},
  journal       = {Physical Review X},
  volume        = {14},
  issue         = {1},
  pages         = {011007},
  numpages      = {18},
  year          = {2024},
  month         = {Jan},
  publisher     = {American Physical Society},
  doi           = {10.1103/PhysRevX.14.011007},
  url           = {https://link.aps.org/doi/10.1103/PhysRevX.14.011007}
}

@article{krantz2019quantum,
  title         = {A quantum engineer's guide to superconducting qubits},
  author        = {Krantz, Philip and Kjaergaard, Morten and Yan, Fei and Orlando, Terry P and Gustavsson, Simon and Oliver, William D},
  journal       = {Applied Physics Reviews},
  volume        = {6},
  number        = {2},
  year          = {2019},
  publisher     = {AIP Publishing},
  url           = {https://pubs.aip.org/aip/apr/article/6/2/021318/570326},
  doi           = {10.1063/1.5089550}
}

@article{groszkowski2018coherence,
  title         = {Coherence properties of the 0-$\pi$ qubit},
  author        = {Groszkowski, Peter and Paolo, A Di and Grimsmo, AL and Blais, A and Schuster, DI and Houck, Andrew Addison and Koch, Jens},
  journal       = {New Journal of Physics},
  volume        = {20},
  number        = {4},
  pages         = {043053},
  year          = {2018},
  publisher     = {IOP Publishing},
  doi           = {10.1088/1367-2630/aab7cd}
}

@article{krojer2024fast,
  title         = {Fast universal control of a flux qubit via exponentially tunable wave-function overlap},
  author        = {Kr\o{}jer, Svend and Dahl, Anders Enevold and Christensen, Kasper Sangild and Kjaergaard, Morten and Flensberg, Karsten},
  journal       = {Physical Review Research},
  volume        = {6},
  issue         = {2},
  pages         = {023064},
  numpages      = {12},
  year          = {2024},
  month         = {Apr},
  publisher     = {American Physical Society},
  doi           = {10.1103/PhysRevResearch.6.023064},
  url           = {https://link.aps.org/doi/10.1103/PhysRevResearch.6.023064}
}

@article{van2006supercurrent,
  title         = {Supercurrent reversal in quantum dots},
  author        = {Van Dam, Jorden A and Nazarov, Yuli V and Bakkers, Erik PAM and De Franceschi, Silvano and Kouwenhoven, Leo P},
  journal       = {Nature},
  volume        = {442},
  number        = {7103},
  pages         = {667--670},
  year          = {2006},
  publisher     = {Nature Publishing Group UK London},
  url           = {https://www.nature.com/articles/nature05018},
  doi           = {10.1038/nature05018}
}

@article{Park_2017,
  title         = {{Andreev spin qubits in multichannel Rashba nanowires}},
  volume        = {96},
  ISSN          = {2469-9969},
  url           = {https://doi.org/10.1103/PhysRevB.96.125416},
  DOI           = {10.1103/physrevb.96.125416},
  number        = {12},
  journal       = {Physical Review B},
  publisher     = {American Physical Society (APS)},
  author        = {Park, Sunghun and Yeyati, A. Levy},
  year          = {2017},
  month         = sep
}

@article{han2023variable,
  title         = {{Variable and orbital-dependent spin-orbit field orientations in an InSb double quantum dot characterized via dispersive gate sensing}},
  author        = {Han, Lin and Chan, Michael and De Jong, Damaz and Prosko, Christian and Badawy, Ghada and Gazibegovic, Sasa and Bakkers, Erik PAM and Kouwenhoven, Leo P and Malinowski, Filip K and Pfaff, Wolfgang},
  journal       = {Physical Review Applied},
  volume        = {19},
  number        = {1},
  pages         = {014063},
  year          = {2023},
  publisher     = {APS},
  url           = {https://journals.aps.org/prapplied/abstract/10.1103/PhysRevApplied.19.014063},
  doi           = {10.1103/PhysRevApplied.19.014063}
}

@article{pita2023direct,
  title         = {Direct manipulation of a superconducting spin qubit strongly coupled to a transmon qubit},
  author        = {Pita-Vidal, Marta and Bargerbos, Arno and {\v{Z}}itko, Rok and Splitthoff, Lukas J and Gr{\"u}nhaupt, Lukas and Wesdorp, Jaap J and Liu, Yu and Kouwenhoven, Leo P and Aguado, Ram{\'o}n and van Heck, Bernard and others},
  journal       = {Nature Physics},
  volume        = {19},
  number        = {8},
  pages         = {1110--1115},
  year          = {2023},
  publisher     = {Nature Publishing Group UK London},
  url           = {https://www.nature.com/articles/s41567-023-02071-x},
  doi           = {10.1038/s41567-023-02071-x}
}

@article{bargerbos2023spectroscopy,
  title         = {{Spectroscopy of spin-split Andreev levels in a quantum dot with superconducting leads}},
  author        = {Bargerbos, Arno and Pita-Vidal, Marta and {\v{Z}}itko, Rok and Splitthoff, Lukas J and Gr{\"u}nhaupt, Lukas and Wesdorp, Jaap J and Liu, Yu and Kouwenhoven, Leo P and Aguado, Ram{\'o}n and Andersen, Christian Kraglund and others},
  journal       = {Physical Review Letters},
  volume        = {131},
  number        = {9},
  pages         = {097001},
  year          = {2023},
  publisher     = {APS},
  url           = {https://journals.aps.org/prl/abstract/10.1103/PhysRevLett.131.097001},
  doi           = {10.1103/PhysRevLett.131.097001}
}

@article{kalashnikov2020bifluxon,
  title         = {{Bifluxon: Fluxon-parity-protected superconducting qubit}},
  author        = {Kalashnikov, Konstantin and Hsieh, Wen Ting and Zhang, Wenyuan and Lu, Wen-Sen and Kamenov, Plamen and Di Paolo, Agustin and Blais, Alexandre and Gershenson, Michael E and Bell, Matthew},
  journal       = {PRX Quantum},
  volume        = {1},
  number        = {1},
  pages         = {010307},
  year          = {2020},
  publisher     = {APS},
  url           = {https://journals.aps.org/prxquantum/abstract/10.1103/PRXQuantum.1.010307},
  doi           = {10.1103/PRXQuantum.1.010307}
}

@article{brooks2013protected,
  title         = {Protected gates for superconducting qubits},
  author        = {Brooks, Peter and Kitaev, Alexei and Preskill, John},
  journal       = {Physical Review A—Atomic, Molecular, and Optical Physics},
  volume        = {87},
  number        = {5},
  pages         = {052306},
  year          = {2013},
  publisher     = {APS},
  url           = {https://journals.aps.org/pra/abstract/10.1103/PhysRevA.87.052306},
  doi           = {10.1103/PhysRevA.87.052306}
}

@article{bell2014protected,
  title         = {{Protected Josephson rhombus chains}},
  author        = {Bell, Matthew T and Paramanandam, Joshua and Ioffe, Lev B and Gershenson, Michael E},
  journal       = {Physical Review Letters},
  volume        = {112},
  number        = {16},
  pages         = {167001},
  year          = {2014},
  publisher     = {APS},
  url           = {https://journals.aps.org/prl/abstract/10.1103/PhysRevLett.112.167001},
  doi           = {10.1103/PhysRevLett.112.167001}
}

@article{smith2020superconducting,
  title         = {{Superconducting circuit protected by two-Cooper-pair tunneling}},
  author        = {Smith, WC and Kou, A and Xiao, X and Vool, U and Devoret, MH},
  journal       = {npj Quantum Information},
  volume        = {6},
  number        = {1},
  pages         = {8},
  year          = {2020},
  publisher     = {Nature Publishing Group UK London},
  url           = {https://www.nature.com/articles/s41534-019-0231-2},
  doi           = {10.1038/s41534-019-0231-2}
}

@article{Padurariu_2010,
  title         = {Theoretical proposal for superconducting spin qubits},
  volume        = {81},
  ISSN          = {1550-235X},
  url           = {https://doi.org/10.1103/PhysRevB.81.144519},
  DOI           = {10.1103/physrevb.81.144519},
  number        = {14},
  journal       = {Physical Review B},
  publisher     = {American Physical Society (APS)},
  author        = {Padurariu, C. and Nazarov, Yu. V.},
  year          = {2010},
  month         = apr
}

@article{Hays_2021,
  title         = {{Coherent manipulation of an Andreev spin qubit}},
  volume        = {373},
  ISSN          = {1095-9203},
  url           = {https://doi.org/10.1126/science.abf0345},
  DOI           = {10.1126/science.abf0345},
  number        = {6553},
  journal       = {Science},
  publisher     = {American Association for the Advancement of Science (AAAS)},
  author        = {Hays, M. and Fatemi, V. and Bouman, D. and Cerrillo, J. and Diamond, S. and Serniak, K. and Connolly, T. and Krogstrup, P. and Nygård, J. and Levy Yeyati, A. and Geresdi, A. and Devoret, M. H.},
  year          = {2021},
  month         = jul,
  pages         = {430–433}
}

@article{Hays_2020,
  title         = {Continuous monitoring of a trapped superconducting spin},
  volume        = {16},
  ISSN          = {1745-2481},
  url           = {https://doi.org/10.1038/s41567-020-0952-3},
  DOI           = {10.1038/s41567-020-0952-3},
  number        = {11},
  journal       = {Nature Physics},
  publisher     = {Springer Science and Business Media LLC},
  author        = {Hays, M. and Fatemi, V. and Serniak, K. and Bouman, D. and Diamond, S. and de Lange, G. and Krogstrup, P. and Nygård, J. and Geresdi, A. and Devoret, M. H.},
  year          = {2020},
  month         = jul,
  pages         = {1103–1107}
}

@article{Pita_Vidal_2024,
  title         = {Strong tunable coupling between two distant superconducting spin qubits},
  volume        = {20},
  ISSN          = {1745-2481},
  url           = {https://doi.org/10.1038/s41567-024-02497-x},
  DOI           = {10.1038/s41567-024-02497-x},
  number        = {7},
  journal       = {Nature Physics},
  publisher     = {Springer Science and Business Media LLC},
  author        = {Pita-Vidal, Marta and Wesdorp, Jaap J. and Splitthoff, Lukas J. and Bargerbos, Arno and Liu, Yu and Kouwenhoven, Leo P. and Andersen, Christian Kraglund},
  year          = {2024},
  month         = may,
  pages         = {1158–1163}
}

@article{Lu_2025,
  title         = {{Andreev spin relaxation time in a shadow-evaporated InAs weak link}},
  volume        = {24},
  ISSN          = {2331-7019},
  url           = {https://doi.org/10.1103/v3lq-t5z8},
  DOI           = {10.1103/v3lq-t5z8},
  number        = {2},
  journal       = {Physical Review Applied},
  publisher     = {American Physical Society (APS)},
  author        = {Lu, Haoran and Bofill, David F. and Sun, Zhenhai and Kanne, Thomas and Nygård, Jesper and Kjaergaard, Morten and Fatemi, Valla},
  year          = {2025},
  month         = aug
}

@article{PhysRevB.109.155164,
  title         = {{Generalized transmon Hamiltonian for Andreev spin qubits}},
  author        = {Pave\ifmmode \check{s}\else \v{s}\fi{}i\ifmmode \acute{c}\else \'{c}\fi{}, Luka and \ifmmode \check{Z}\else \v{Z}\fi{}itko, Rok},
  journal       = {Physical Review B},
  volume        = {109},
  issue         = {15},
  pages         = {155164},
  numpages      = {23},
  year          = {2024},
  month         = {Apr},
  publisher     = {American Physical Society},
  doi           = {10.1103/PhysRevB.109.155164},
  url           = {https://link.aps.org/doi/10.1103/PhysRevB.109.155164}
}

@article{PhysRevLett.90.226806,
  title         = {Andreev Quantum Dots for Spin Manipulation},
  author        = {Chtchelkatchev, Nikolai M. and Nazarov, Yu. V.},
  journal       = {Physical Review Letters},
  volume        = {90},
  issue         = {22},
  pages         = {226806},
  numpages      = {4},
  year          = {2003},
  month         = {Jun},
  publisher     = {American Physical Society},
  doi           = {10.1103/PhysRevLett.90.226806},
  url           = {https://link.aps.org/doi/10.1103/PhysRevLett.90.226806}
}

@misc{manesco2026looplessmultiterminalquantumcircuits,
  title         = {Loopless multiterminal quantum circuits at odd parity},
  author        = {Antonio Manesco and Anton Akhmerov and Valla Fatemi},
  year          = {2026},
  eprint        = {2601.13369},
  archivePrefix = {arXiv},
  primaryClass  = {cond-mat.mes-hall},
  url           = {https://arxiv.org/abs/2601.13369},
  doi           = {10.48550/arXiv.2601.13369}
}

@article{Liu_2019,
  title         = {{Semiconductor–Ferromagnetic Insulator–Superconductor Nanowires: Stray Field and Exchange Field}},
  volume        = {20},
  ISSN          = {1530-6992},
  url           = {https://doi.org/10.1021/acs.nanolett.9b04187},
  DOI           = {10.1021/acs.nanolett.9b04187},
  number        = {1},
  journal       = {Nano Letters},
  publisher     = {American Chemical Society (ACS)},
  author        = {Liu, Yu and Vaitiekėnas, Saulius and Martí-Sánchez, Sara and Koch, Christian and Hart, Sean and Cui, Zheng and Kanne, Thomas and Khan, Sabbir A. and Tanta, Rawa and Upadhyay, Shivendra and Cachaza, Martin Espiñeira and Marcus, Charles M. and Arbiol, Jordi and Moler, Kathryn A. and Krogstrup, Peter},
  year          = {2019},
  month         = nov,
  pages         = {456–462}
}

@article{PhysRevB.105.L041304,
  title         = {Evidence for spin-polarized bound states in semiconductor--superconductor--ferromagnetic-insulator islands},
  author        = {Vaitiek\ifmmode \dot{e}\else \.{e}\fi{}nas, S. and Souto, R. Seoane and Liu, Y. and Krogstrup, P. and Flensberg, K. and Leijnse, M. and Marcus, C. M.},
  journal       = {Physical Review B},
  volume        = {105},
  issue         = {4},
  pages         = {L041304},
  numpages      = {5},
  year          = {2022},
  month         = {Jan},
  publisher     = {American Physical Society},
  doi           = {10.1103/PhysRevB.105.L041304},
  url           = {https://link.aps.org/doi/10.1103/PhysRevB.105.L041304}
}

@article{Jiang_2025,
  title         = {Zero-bias conductance peaks at zero applied magnetic field due to stray fields from integrated micromagnets in hybrid nanowire quantum dots},
  volume        = {19},
  ISSN          = {2542-4653},
  url           = {https://doi.org/10.21468/SciPostPhys.19.2.030},
  DOI           = {10.21468/scipostphys.19.2.030},
  number        = {2},
  journal       = {SciPost Physics},
  publisher     = {Stichting SciPost},
  author        = {Jiang, Luyao and Gupta, Mohit and Riggert, C. and Pendharkar, M. and Dempsey, C. and Lee, Sungjay and Harrington, S. D. and Palmstrøm, C. J. and Pribiag, V. S. and Frolov, Sergey M.},
  year          = {2025},
  month         = aug
}

@article{Vaitiek_nas_2020,
  title         = {Zero-bias peaks at zero magnetic field in ferromagnetic hybrid nanowires},
  volume        = {17},
  ISSN          = {1745-2481},
  url           = {https://doi.org/10.1038/s41567-020-1017-3},
  DOI           = {10.1038/s41567-020-1017-3},
  number        = {1},
  journal       = {Nature Physics},
  publisher     = {Springer Science and Business Media LLC},
  author        = {Vaitiekėnas, S. and Liu, Y. and Krogstrup, P. and Marcus, C. M.},
  year          = {2020},
  month         = sep,
  pages         = {43–47}
}

@article{Lakic_2025,
  title         = {A quantum dot in germanium proximitized by a superconductor},
  volume        = {24},
  ISSN          = {1476-4660},
  url           = {https://doi.org/10.1038/s41563-024-02095-5},
  DOI           = {10.1038/s41563-024-02095-5},
  number        = {4},
  journal       = {Nature Materials},
  publisher     = {Springer Science and Business Media LLC},
  author        = {Lakic, Lazar and Lawrie, William I. L. and van Driel, David and Stehouwer, Lucas E. A. and Su, Yao and Veldhorst, Menno and Scappucci, Giordano and Kuemmeth, Ferdinand and Chatterjee, Anasua},
  year          = {2025},
  month         = feb,
  pages         = {552–558}
}

@misc{fabris2026granularaluminuminducedsuperconductivity,
  title         = {Granular aluminum induced superconductivity in germanium for hole spin-based hybrid devices},
  author        = {Giorgio Fabris and Paul Falthansl-Scheinecker and Devashish Shah and Daniel Michel Pino and Maksim Borovkov and Anton Bubis and Kevin Roux and Dina Sokolova and Alejandro Andres Juanes and Tommaso Costanzo and Inas Taha and Aziz Genç and Jordi Arbiol and Stefano Calcaterra and Afonso De Cerdeira Oliveira and Daniel Chrastina and Giovanni Isella and Ruben Seoane Souto and Maria Jose Calderon and Ramon Aguado and Jose Carlos Abadillo-Uriel and Georgios Katsaros},
  year          = {2026},
  eprint        = {2602.21364},
  archivePrefix = {arXiv},
  primaryClass  = {cond-mat.mes-hall},
  url           = {https://arxiv.org/abs/2602.21364},
  doi           = {10.48550/arXiv.2602.21364}
}

@article{PRXQuantum.5.030357,
  title         = {{Direct Microwave Spectroscopy of Andreev Bound States in Planar $\mathrm{Ge}$ Josephson Junctions}},
  author        = {Hinderling, M. and ten Kate, S. C. and Coraiola, M. and Haxell, D.Z. and Stiefel, M. and Mergenthaler, M. and Paredes, S. and Bedell, S.W. and Sabonis, D. and Nichele, F.},
  journal       = {PRX Quantum},
  volume        = {5},
  issue         = {3},
  pages         = {030357},
  numpages      = {14},
  year          = {2024},
  month         = {Sep},
  publisher     = {American Physical Society},
  doi           = {10.1103/PRXQuantum.5.030357},
  url           = {https://link.aps.org/doi/10.1103/PRXQuantum.5.030357}
}

@article{jung2025tunneling,
  title         = {{Tunneling spectroscopy of Andreev bands in multiterminal graphene-based Josephson junctions}},
  author        = {Jung, Woochan and Jin, Seyoung and Park, Sein and Shin, Seung-Hyun and Watanabe, Kenji and Taniguchi, Takashi and Cho, Gil Young and Lee, Gil-Ho},
  journal       = {Science Advances},
  volume        = {11},
  number        = {21},
  pages         = {eads0342},
  year          = {2025},
  publisher     = {American Association for the Advancement of Science},
  doi           = {10.1126/sciadv.ads0342}
}

@article{Riechert_2025,
  title         = {The carbon nanotube gatemon qubit},
  volume        = {16},
  ISSN          = {2041-1723},
  url           = {https://doi.org/10.1038/s41467-025-62283-y},
  DOI           = {10.1038/s41467-025-62283-y},
  number        = {1},
  journal       = {Nature Communications},
  publisher     = {Springer Science and Business Media LLC},
  author        = {Riechert, H. and Annabi, S. and Peugeot, A. and Duprez, H. and Hantute, M. and Watanabe, K. and Taniguchi, T. and Arrighi, E. and Griesmar, J. and Pillet, J.-D. and Bretheau, L.},
  year          = {2025},
  month         = aug
}

@article{Chakraborty_2025,
  title         = {Cross-platform protected qubits from entanglement},
  volume        = {112},
  ISSN          = {2469-9969},
  url           = {https://doi.org/10.1103/qjhp-8x6z},
  DOI           = {10.1103/qjhp-8x6z},
  number        = {15},
  journal       = {Physical Review B},
  publisher     = {American Physical Society (APS)},
  author        = {Chakraborty, Nilotpal and Moessner, Roderich and Doucot, Benoit},
  year          = {2025},
  month         = oct
}

@article{PRXQuantum.2.030101,
  title         = {{Moving beyond the Transmon: Noise-Protected Superconducting Quantum Circuits}},
  author        = {Gyenis, Andr\'as and Di Paolo, Agustin and Koch, Jens and Blais, Alexandre and Houck, Andrew A. and Schuster, David I.},
  journal       = {PRX Quantum},
  volume        = {2},
  issue         = {3},
  pages         = {030101},
  numpages      = {15},
  year          = {2021},
  month         = {Sep},
  publisher     = {American Physical Society},
  doi           = {10.1103/PRXQuantum.2.030101},
  url           = {https://link.aps.org/doi/10.1103/PRXQuantum.2.030101}
}

@article{Danon_2021,
  title         = {Protected solid-state qubits},
  volume        = {119},
  ISSN          = {1077-3118},
  url           = {https://doi.org/10.1063/5.0073945},
  DOI           = {10.1063/5.0073945},
  number        = {26},
  journal       = {Applied Physics Letters},
  publisher     = {AIP Publishing},
  author        = {Danon, Jeroen and Chatterjee, Anasua and Gyenis, András and Kuemmeth, Ferdinand},
  year          = {2021},
  month         = dec
}

@article{Matute_Ca_adas_2024,
  title         = {{Quantum Circuits with Multiterminal Josephson-Andreev Junctions}},
  volume        = {5},
  ISSN          = {2691-3399},
  url           = {https://doi.org/10.1103/PRXQuantum.5.020340},
  DOI           = {10.1103/prxquantum.5.020340},
  number        = {2},
  journal       = {PRX Quantum},
  publisher     = {American Physical Society (APS)},
  author        = {Matute-Cañadas, F.J. and Tosi, L. and Yeyati, A. Levy},
  year          = {2024},
  month         = may
}

@misc{caceres2026ferbonoiseresilientqubit,
  title         = {{FerBo: a noise resilient qubit hybridizing Andreev and fluxonium states}},
  author        = {J. J. Caceres and D. Sanz Marco and J. Ortuzar and E. Flurin and C. Urbina and H. Pothier and M. F. Goffman and F. J. Matute-Cañadas and A. Levy Yeyati},
  year          = {2026},
  eprint        = {2604.01145},
  archivePrefix = {arXiv},
  primaryClass  = {cond-mat.mes-hall},
  url           = {https://arxiv.org/abs/2604.01145},
  doi           = {10.48550/arXiv.2604.01145}
}

@article{Cochrane_1999,
  title         = {Macroscopically distinct quantum-superposition states as a bosonic code for amplitude damping},
  volume        = {59},
  ISSN          = {1094-1622},
  url           = {https://doi.org/10.1103/PhysRevA.59.2631},
  DOI           = {10.1103/physreva.59.2631},
  number        = {4},
  journal       = {Physical Review A},
  publisher     = {American Physical Society (APS)},
  author        = {Cochrane, P. T. and Milburn, G. J. and Munro, W. J.},
  year          = {1999},
  month         = apr,
  pages         = {2631–2634}
}

@article{Mirrahimi_2014,
  title         = {Dynamically protected cat-qubits: a new paradigm for universal quantum computation},
  volume        = {16},
  ISSN          = {1367-2630},
  url           = {https://doi.org/10.1088/1367-2630/16/4/045014},
  DOI           = {10.1088/1367-2630/16/4/045014},
  number        = {4},
  journal       = {New Journal of Physics},
  publisher     = {IOP Publishing},
  author        = {Mirrahimi, Mazyar and Leghtas, Zaki and Albert, Victor V and Touzard, Steven and Schoelkopf, Robert J and Jiang, Liang and Devoret, Michel H},
  year          = {2014},
  month         = apr,
  pages         = {045014}
}

@misc{roverch2026experimentalrealizationcos2varphitransmon,
  title         = {Experimental realization of a $\cos(2\varphi)$ transmon qubit},
  author        = {Erwan Roverc'h and Alvise Borgognoni and Marius Villiers and Kyrylo Gerashchenko and W. Clarke Smith and Christopher Wilson and Benoit Douçot and Alexandru Petrescu and Philippe Campagne-Ibarcq and Zaki Leghtas},
  year          = {2026},
  eprint        = {2603.13114},
  archivePrefix = {arXiv},
  primaryClass  = {quant-ph},
  url           = {https://arxiv.org/abs/2603.13114},
  doi           = {10.48550/arXiv.2603.13114}
}

@article{Ioffe_2002,
  title         = {Possible realization of an ideal quantum computer in Josephson junction array},
  volume        = {66},
  ISSN          = {1095-3795},
  url           = {https://doi.org/10.1103/PhysRevB.66.224503},
  DOI           = {10.1103/physrevb.66.224503},
  number        = {22},
  journal       = {Physical Review B},
  publisher     = {American Physical Society (APS)},
  author        = {Ioffe, L. B. and Feigel’man, M. V.},
  year          = {2002},
  month         = dec
}

@article{Bao_2022,
  title         = {{Fluxonium: An Alternative Qubit Platform for High-Fidelity Operations}},
  volume        = {129},
  ISSN          = {1079-7114},
  url           = {https://doi.org/10.1103/PhysRevLett.129.010502},
  DOI           = {10.1103/physrevlett.129.010502},
  number        = {1},
  journal       = {Physical Review Letters},
  publisher     = {American Physical Society (APS)},
  author        = {Bao, Feng and Deng, Hao and Ding, Dawei and Gao, Ran and Gao, Xun and Huang, Cupjin and Jiang, Xun and Ku, Hsiang-Sheng and Li, Zhisheng and Ma, Xizheng and Ni, Xiaotong and Qin, Jin and Song, Zhijun and Sun, Hantao and Tang, Chengchun and Wang, Tenghui and Wu, Feng and Xia, Tian and Yu, Wenlong and Zhang, Fang and Zhang, Gengyan and Zhang, Xiaohang and Zhou, Jingwei and Zhu, Xing and Shi, Yaoyun and Chen, Jianxin and Zhao, Hui-Hai and Deng, Chunqing},
  year          = {2022},
  month         = jun
}

@misc{fu2026fluxnoiseresilienttransmonqubitdoublyconnected,
  title         = {Flux-noise-resilient transmon qubit via a doubly-connected gradiometric design},
  author        = {J. B. Fu and Da-Wei Wang and B. Ren and Z. H. Yang and S. Hu and G. Y. Huang and S. H. Cao and D. D. Liu and X. F. Zhang and X. Fu and S. C. Xue and Y. G. Che and Yu-xi Liu and M. T. Deng and J. J. Wu},
  year          = {2026},
  eprint        = {2601.02137},
  archivePrefix = {arXiv},
  primaryClass  = {quant-ph},
  url           = {https://arxiv.org/abs/2601.02137},
  doi           = {10.48550/arXiv.2601.02137}
}

@article{Schwarz_2013,
  title         = {Gradiometric flux qubits with a tunable gap},
  volume        = {15},
  ISSN          = {1367-2630},
  url           = {https://doi.org/10.1088/1367-2630/15/4/045001},
  DOI           = {10.1088/1367-2630/15/4/045001},
  number        = {4},
  journal       = {New Journal of Physics},
  publisher     = {IOP Publishing},
  author        = {Schwarz, M J and Goetz, J and Jiang, Z and Niemczyk, T and Deppe, F and Marx, A and Gross, R},
  year          = {2013},
  month         = apr,
  pages         = {045001}
}

@article{bulaevskii1977superconducting,
  title         = {Superconducting system with weak coupling to the current in the ground state},
  author        = {Bulaevskii, LN and Kuzii, VV and Sobyanin, AA},
  journal       = {JETP Letters},
  volume        = {25},
  number        = {7},
  pages         = {290--294},
  year          = {1977}
}

@article{hutchings2017tunable,
  title         = {Tunable superconducting qubits with flux-independent coherence},
  author        = {Hutchings, MD and Hertzberg, Jared B and Liu, Yebin and Bronn, Nicholas T and Keefe, George A and Brink, Markus and Chow, Jerry M and Plourde, BLT},
  journal       = {Physical Review Applied},
  volume        = {8},
  number        = {4},
  pages         = {044003},
  year          = {2017},
  publisher     = {APS},
  doi           = {10.1103/PhysRevApplied.8.044003}
}

@article{zorin1996background,
  title         = {Background charge noise in metallic single-electron tunneling devices},
  author        = {Zorin, AB and Ahlers, F-J and Niemeyer, J and Weimann, Th and Wolf, H and Krupenin, VA and Lotkhov, SV},
  journal       = {Physical Review B},
  volume        = {53},
  number        = {20},
  pages         = {13682},
  year          = {1996},
  publisher     = {APS},
  doi           = {10.1103/PhysRevB.53.13682}
}

@misc{kurilovich2025highfrequencyreadoutfreetransmon,
  title         = {High-frequency readout free from transmon multi-excitation resonances},
  author        = {Pavel D. Kurilovich and Thomas Connolly and Charlotte G. L. Bøttcher and Daniel K. Weiss and Sumeru Hazra and Vidul R. Joshi and Andy Z. Ding and Heekun Nho and Spencer Diamond and Vladislav D. Kurilovich and Wei Dai and Valla Fatemi and Luigi Frunzio and Leonid I. Glazman and Michel H. Devoret},
  year          = {2025},
  eprint        = {2501.09161},
  archivePrefix = {arXiv},
  primaryClass  = {quant-ph},
  url           = {https://arxiv.org/abs/2501.09161},
  doi           = {10.48550/arXiv.2501.09161}
}

@dataset{tjernshaugen_2026_20082243,
  author        = {Tjernshaugen, Johanne Bratland and
                  Viñas Boström, Florinda and
                  Danon, Jeroen and
                  Linder, Jacob and
                  Flensberg, Karsten and
                  Rigotti Manesco, Antonio Lucas},
  title         = {Coherent control of spinmons},
  month         = may,
  year          = 2026,
  publisher     = {Zenodo},
  doi           = {10.5281/zenodo.20082243},
  url           = {https://doi.org/10.5281/zenodo.20082243}
}

@article{purkayastha2026transmon,
  title={{Transmon qubit using Sn as a junction superconductor}},
  author={Purkayastha, Amrita and Sharma, Amritesh and Patel, Param J and Chen, An-Hsi and Dempsey, Connor P and Asodekar, Shreyas and Sinha, Subhayan and Tomasian, Maxime and Pendharkar, Mihir and Palmstr{\o}m, Christopher J and others},
  journal={Nano Letters},
  volume={26},
  number={6},
  pages={2074--2079},
  year={2026},
  publisher={ACS Publications},
  doi = {10.1021/acs.nanolett.5c05276}
}

@article{isakov2026observation,
  title={Observation of coherent flux-charge interaction in a gate-tunable fluxonium},
  author={Isakov, Brian D and Singh, Shikhar and Parra-Rodriguez, Adrian and Feldstein-Bofill, David and Sun, Zhenhai and Kringh{\o}j, Anders and Kr{\o}jer, Svend and Blais, Alexandre and Kjaergaard, Morten and Gyenis, Andr{\'a}s},
  journal={arXiv preprint arXiv:2607.07798},
  year={2026},
  doi={10.48550/arXiv.2607.07798
}
}

@article{PhysRevLett.120.100502,
  title = {Evolution of Nanowire Transmon Qubits and Their Coherence in a Magnetic Field},
  author = {Luthi, F. and Stavenga, T. and Enzing, O. W. and Bruno, A. and Dickel, C. and Langford, N. K. and Rol, M. A. and Jespersen, T. S. and Nyg\aa{}rd, J. and Krogstrup, P. and DiCarlo, L.},
  journal = {Phys. Rev. Lett.},
  volume = {120},
  issue = {10},
  pages = {100502},
  numpages = {6},
  year = {2018},
  month = {Mar},
  publisher = {American Physical Society},
  doi = {10.1103/PhysRevLett.120.100502},
  url = {https://link.aps.org/doi/10.1103/PhysRevLett.120.100502}
}

@article{Zatelli_2024,
   title={Robust poor man’s Majorana zero modes using Yu-Shiba-Rusinov states},
   volume={15},
   ISSN={2041-1723},
   url={http://dx.doi.org/10.1038/s41467-024-52066-2},
   DOI={10.1038/s41467-024-52066-2},
   number={1},
   journal={Nature Communications},
   publisher={Springer Science and Business Media LLC},
   author={Zatelli, Francesco and van Driel, David and Xu, Di and Wang, Guanzhong and Liu, Chun-Xiao and Bordin, Alberto and Roovers, Bart and Mazur, Grzegorz P. and van Loo, Nick and Wolff, Jan C. and Bozkurt, A. Mert and Badawy, Ghada and Gazibegovic, Sasa and Bakkers, Erik P. A. M. and Wimmer, Michael and Kouwenhoven, Leo P. and Dvir, Tom},
   year={2024},
   month=Sept }

\end{document}